\documentclass[prb,floatfix,superscriptaddress,twocolumn]{revtex4}
\usepackage{graphicx}
\usepackage{color}
\begin{document}
\title{Dynamics and stability of dispersions of polyelectrolyte-filled
multilayer microcapsules}

\author{Byoung-Suhk Kim}
\affiliation{Sogang University, 1 Shinsu-Dong, Mapo-Gu, Seoul
121-742, South Korea}

\author{Vladimir Lobaskin}
\affiliation{Technische Universit\"at M\"unchen,
James-Franck-Str., D-85747 Garching, Germany}

\affiliation{Chelyabinsk State University, Br. Kashirinykh 129,
454126 Chelyabinsk, Russia}

\author{Olga I. Vinogradova}
\email[Corresponding author: ]{oivinograd@yahoo.com} \affiliation{
A.N.Frumkin Institute of Physical Chemistry and Electrochemistry,
Russian Academy of Sciences, Leninsky Prospect 31, 119991 Moscow,
Russia}
\begin{abstract}
We report dynamic and coagulation properties of a dispersion of
polyelectrolyte multilayer microcapsules filled with solutions of
a strong polyelectrolyte. The capsule self-diffusion coefficient
in the vicinity of the wall is measured using a particle tracking
procedure from confocal images of the dispersion. Our results
suggest that the microcapsules take a charge of encapsulated
polyions, which indicates a semi-permeability of the shell and a
leakage of counter-ions. The diffusion of capsules in the force
field is qualitatively similar to that of charged solid particles:
The effective interaction potential contains a weak local
attractive minimum and an electrostatic barrier. We also found
that the aggregation of suspended capsules  occurs faster than
their sedimentation and adhesion onto a glass surface.

\end{abstract}

\maketitle

\section{Introduction}

In recent years, there has been much interest in studying
polyelectrolyte multilayer microcapsules, mostly
``hollow''~\cite{donath.e:1998,caruso.f:1998}, but also filled
with the neutral or charged polymer solutions. Such ``filled''
microcapsules represent a novel type of nano-engineered composite
microstructures. They can be prepared by a variety of
methods~\cite{radtchenko.il:2000,sukhorukov.gb:2001.b,dahne.l:2001,lulevich.vv:2002,vinogradova.oi:2004}
and are potentially important in many areas of science and
technology. For instance, they allow one to mimic the advanced
systems containing biopolymers and could serve as a new composite
material with controlled
stiffness~\cite{sukhorukov.gb:2004,kim.bs:2004,lebedeva.ov:2004,kim.bs:2005c}.

The potential applications of polyelectrolyte microcapsules to a
big extent depend on how well their physical properties are
understood and could be controlled. So far, main attention was
attracted by mechanical and adhesion
properties~\cite{gao.c:2001,lulevich.vv:2002,lulevich.vv:2003,elsner.n:2004,vinogradova.oi:2004,lebedeva.ov:2004,vinogradova.oi:2006},
since they define deformation and rupture of the capsule shells
under external load, which is important for protection of
encapsulated material in delivery/release systems (for recent
reviews see~\cite{vinogradova.oi:2004b,vinogradova.oi:2006}).
Beside the knowledge about microcapsule mechanical behavior, the
topics of big relevance to numerous potential applications are the
stability and dynamics of colloidal dispersions of microcapsules.
The potential use of capsules as containers in pharmaceutical or
chemical applications strongly depends on the possibility to
prepare their stable dispersions. The questions of interest
include the dynamic and diffusion properties of microcapsule
dispersion; whether and under which conditions this dispersion is
stable or tends to aggregate; how do capsules interact with the
surfaces; and what kinds of long- and short-range interactions are
expected. Although these questions might be considered as central
for any colloidal systems, and despite enormous experimental
activity devoted to design of new functional polyelectrolyte
microcapsules~\cite{shchukin.dg:2003,kim.bs:2005b,vinogradova:oi:2005,kim.bs:2006},
these dynamics and stability issues were never addressed.

The lack of information about interactions and aggregation in the
dispersion of microcapsules is not the least due to the absence of
direct experimental methods to study interactions in the capsule
systems. Thus, the surface force apparatus measures the
interactions between macroscopic bodies which are at least 1000
times larger than
microcapsules~\cite{israelachvili.jn:1978,connor.jn:2001}, not to
say that it requires that at least one interacting surface is
rigid. The use of a single polyelectrolyte microcapsule as a
colloidal probe in the atomic force microscope
approach~\cite{ducker.wa:1991} is hardly possible, since the
standard methods of attachment of a probe to a cantilever would
lead to a destruction/collapse of the capsules. Optical tweezers
techniques~\cite{brunner.m:2004a} normally require a much larger
contrast in dielectric constants. The same remark concerns methods
based on total internal reflection
microscopy~\cite{prieve.dc:1990}, which would only be possible in
case of significant difference in refractive indices. The only
potentially suitable known technique that could be used for
studying the interaction properties of microcapsules is that based
on direct measurements and subsequent inversion of a pair
distribution
function~\cite{kepler.gm:1994,behrens.sh:2001,brunner.m:2002,lobaskin.v:2003a}.
This method (involving numerical image analysis), however, was
used so far only for solid particles. The reconstruction of
capsule positions from the confocal image is straightforward for a
stable dispersion, where the particles are well separated in
space. However, the evaluation of the potential from the images of
an aggregating dispersion is complicated by (i) non-trivial
fluorescence intensity profiles of filled capsules and (ii) by a
substantial overlap of intensity profiles of contacting species.
The calculation of distribution functions would require therefore
measurements of much higher resolution and precision than in case
of solid particles.

In this paper, we probe the interaction potential in a partially
stabilized dispersion of ``filled'' microcapsules suspended in
salt-free water. Our approach is based on the reconstruction of
the potential from the measured coagulation rate and translation
speed of capsules along the plate. To the best of our knowledge,
our paper represents the first attempt to evaluate an interaction
(pair) potential of the capsules. Our consideration here is rather
approximate due to inevitable difficulties caused by a complexity
of a microcapsule's system and, as a consequence, some inaccuracy
of measurements and difficulties in interpretation of experimental
data. However, our approaches provide us with some guidance and
lead to unambiguous conclusions. Thus, we show that ``filled''
capsules take a charge of an encapsulated polyelectrolyte
indicating semi-permeable properties of the multilayer shell and a
counter-ion leakage. The diffusion of ``filled'' capsules in a
force field is similar to that of charged solid particles. We also
prove that the interaction potential contains an electrostatic
barrier with a weak local attractive minimum close to a contact of
capsules. Our results show that encapsulation of charged polymers
is as important in its effect on adhesion of microcapsules and
their long-range interactions as for other properties studied
before.

\section{Experimental}

\subsection{Materials}

The fluorescent dye Rhodamine B isothiocyanate (RBITC),
shell-forming polyelectrolytes poly(sodium 4-styrenesulfonate) (PSS;
Mw $\sim$ 70 kDa) and poly(allylamine hydrochloride) (PAH; Mw $\sim$
70 kDa) were purchased from Sigma-Aldrich Chemie GmbH, Germany.
Sodium chloride (NaCl) was purchased from Riedel-de Haen, Germany.
All chemicals were of analytical purity or higher quality and were
used without further purification. Suspensions of monodisperse
weakly cross-linked melamine formaldehyde particles (MF-particles)
with a radius of 2 $\mu m$ were purchased from Microparticles GmbH
(Berlin, Germany). Glass bottom dishes (0.17 mm/ 30 mm) with optical
quality surfaces were obtained from World Precision Instruments Inc.
(USA).

Fluorescent PSS-RBITC for encapsulation was prepared according to a
method published in Ref. \cite{vinogradova.oi:2004,kim.bs:2004}.
Briefly, labeled allylamine was first produced, and afterwards mixed
with sodium styrenesulfonate (SS), and then copolymerized under
N$_2$ radically. The allylamine was mixed with RBITC dissolved in
ethanol. The mixture was stirred for four hours at room temperature.
Afterwards, SS was added in an amount corresponding to about 200
monomer equivalents. Then as initiator 1\% K$_2$S$_2$O$_8$, related
to the monomer concentration, was added to this solution for a
radical polymerization. The mixture was heated up to 80$^{\circ}$C
and was stirred for four hours in a nitrogen atmosphere. After
polymerization, labeled polymer was dialyzed extensively against
distilled water using 3500 molecular-weight-cutoff dialysis tubing
until no more colour could be observed in the wash water. The
dialyzed polymer solution was freeze-dried. The weight-average
molecular weight (Mw) was estimated by gel-permeation chromatography
(column: TSK Gel G6000, TSK Gel G5000, TSK Gel G3000) with standard
polyethylene oxide as a reference using H$_2$O as an eluent at
23$^{\circ}$C. Mw and polydispersity were found as about 46 kDa and
1.86, respectively. Water used for all experiments was purified by a
commercial Milli-Q Gradient A10 system containing ion exchange and
charcoal stages, and had a high resistivity of 18.2 M$\Omega$/cm.

\subsection{Methods}

\subsubsection{Capsule preparation}

The positively charged MF particles (50  mL of 10 wt \%
dispersion) as a template were incubated with 1 mL of the
negatively charged PSS solution (1 mg/mL containing 0.5 mol/L
NaCl, pH 6) at room temperature for 10 min, followed by three
centrifugation/rinsing cycles, and finally dispersed in water. 1
mL of a PAH solution (1 mg/mL containing 0.5 mol/L, pH 6) was then
added to the particle dispersion. After 10 min given for
adsorption three centrifugation/wash cycles were performed (as
above). The PSS and PAH adsorption steps were repeated four times
each to build multilayers on the MF particles. Washing out excess
polymer and salt were followed by each adsorption. The
microcapsules referred to below as ``hollow'' capsules were
obtained by dissolving the MF template in HCl at pH 1.2-1.6 and
washing with water three times as described
before~\cite{sukhorukov.gb:1998.b}.

The ``hollow'' capsules were then filled with polyelectrolyte. The
detailed procedures are described in the previous
report.~\cite{vinogradova.oi:2004} The encapsulation of
polyelectrolyte included several steps. Briefly, the original
``hollow'' capsules were exposed to acetone/water mixture (1:1) to
make a multilayer shell permeable for a high molecular weight
polymer~\cite{lulevich.vv:2003} and then RBITC-labeled PSS molecules
were added to the mixtures. During the encapsulation process the PSS
concentration was increased gradually to avoid an osmotic collapse
of the microcapsules~\cite{vinogradova.oi:2004,kim.bs:2004} The
initial PSS concentration was 2 mg/mL ($\sim 0.01$ mol/L) and was
doubled every hour. When the required concentration was reached, the
mixture was diluted with pure water and the multilayer shells were
assumed to return to an impermeable state. The ``filled'' capsules
were separated from the PSS in the bulk solution by centrifugation.
Afterwards, washing cycles with pure water were carried out at least
twice to remove the excess PSS molecules.

\subsubsection{Confocal microscopy}

\begin{figure}
\begin{center}
\vskip 0.2in
\includegraphics[width=8cm]{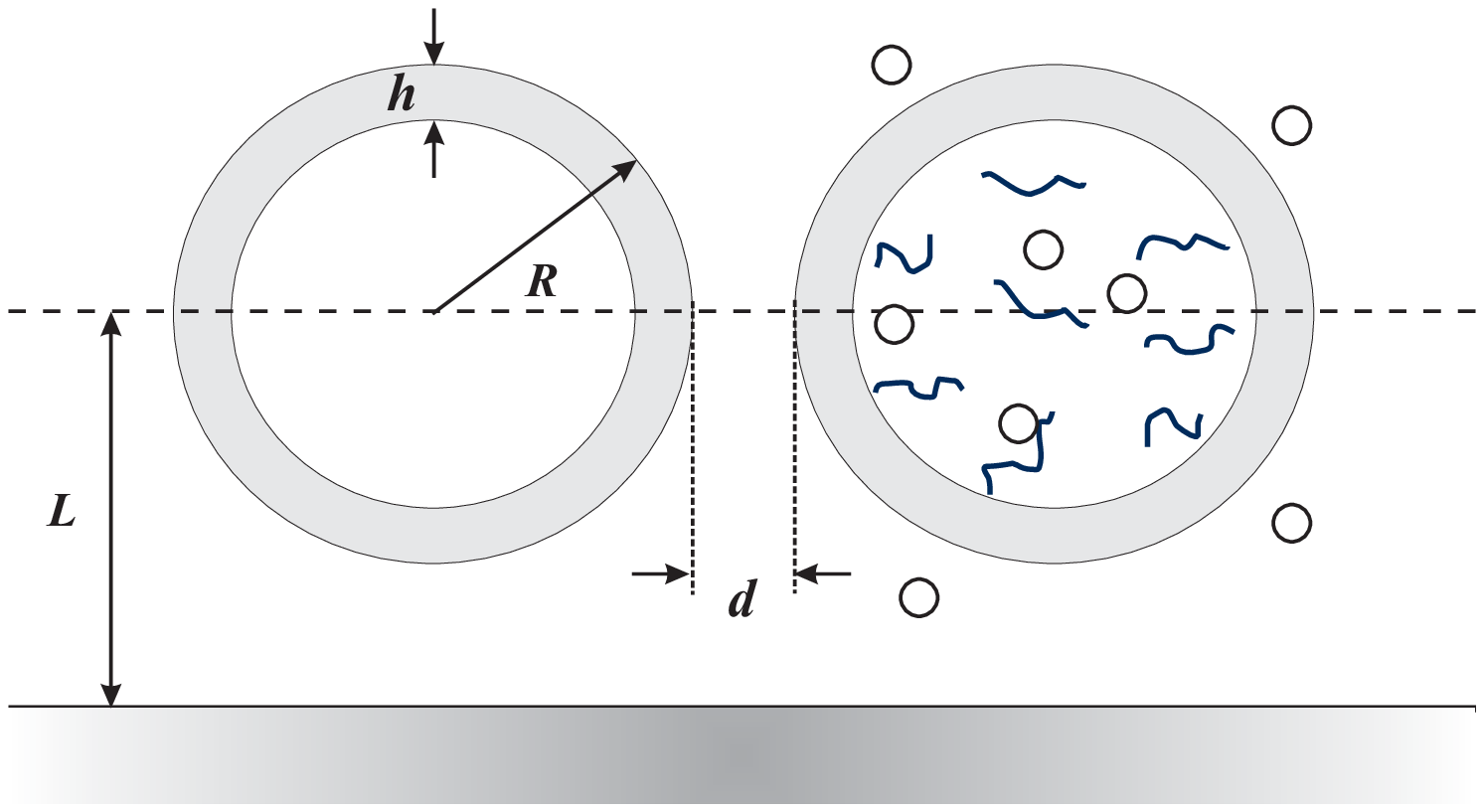}\\
\includegraphics[width=3.8cm]{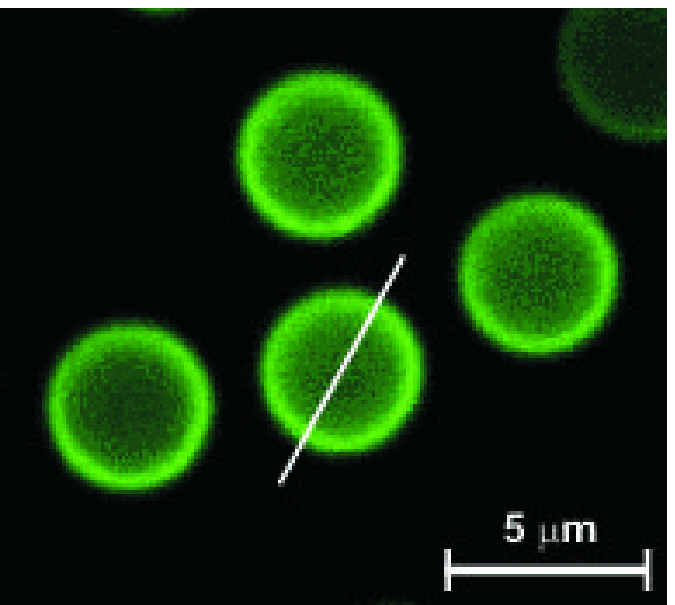}
\includegraphics[width=4.5cm]{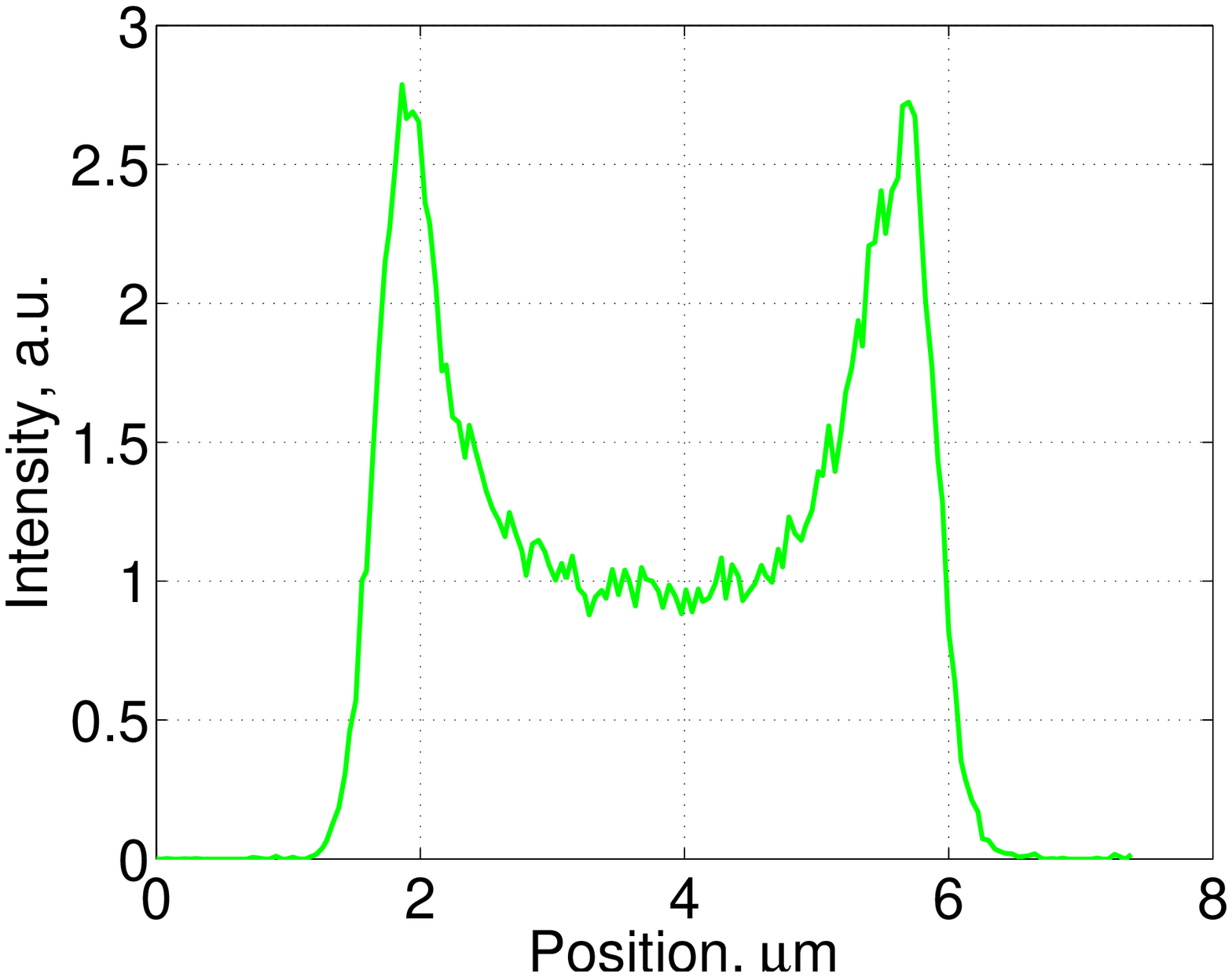}
\end{center}
\caption{Top: Schematic of location of capsules in the vicinity of
the similarly charged wall and the main notations. Polymers and
counter-ions are not shown (left). The illustration of counter-ion
leakage. The charge of the capsule is no longer equal to that of
the shell and is controlled by the charge of inner polymers
(right). Bottom: Confocal image of isolated immobilized capsules
(left) and a typical distribution of encapsulated polyelectrolyte
inside them (right) reflected by measured intensity of the
fluorescence signal.
         }
\label{fig:exam1_a}
\end{figure}

To scan the dynamic motion of ``filled'' microcapsules in the
plane located close to a glass wall (see Fig.~\ref{fig:exam1_a},
top), we used a commercial confocal microscope manufactured by
Olympus (Japan) consisting of the confocal scanning unit Olympus
FV 300 in combination with an inverted microscope Olympus IX70
equipped with a high resolution $60\times$ oil (N.A. 1.45)
immersion objective. The excitation wavelength was chosen
according to the label Rhodamine ($\lambda = 543$ nm).
Microcapsules were dispersed in salt-free water. Then confocal
images of their suspension in the vicinity of the wall were taken
with time interval of 1 s. We normally had about 150 capsules in
the field of view $100 \times 134 \mu m$. Then, the particle
tracking was performed using the IDL software and tracking scripts
by Crocker and Grier \cite{crocker.jc:1996}. As a result, the
two-dimensional mean-square displacements of the center-of-mass of
``filled'' microcapsules were measured. The apparent diffusion
coefficient was estimated either from the jump length distribution
or from the mean-square displacement of the ``filled''
microcapsules as a function of time.

The same confocal setup was used to scan the capsule shape and to
check the concentration distribution of PSS inside the capsules.
These measurements were done after the sedimentation of capsules
with only those of them that remained isolated (see
Fig.~\ref{fig:exam1_a}, bottom) The z-position scanning was done
in steps of 0.02-0.05 $\mu$m. The diameters of the capsules were
determined optically with an accuracy of 0.4 $\mu$m. Concentration
measurements were performed via the fluorescence intensity coming
from the interior of the PSS containing capsules.

\section{Results and discussion}

\subsection{Main observations}

\begin{figure}
\begin{center}
\vskip 0.2in
\includegraphics[width=8cm]{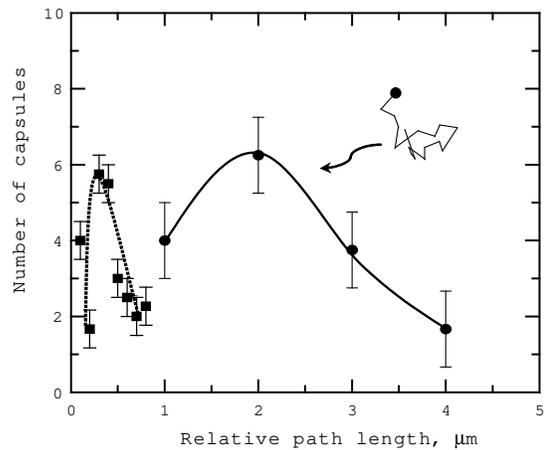}
\end{center}
\caption{Typical trajectories (cartoon) and jump length
distributions of the polyelectrolyte-filled capsules capsules
taken one day (circles) and four days (squares) after the
preparation.
         }
\label{fig:1}
\end{figure}

\begin{figure}
\begin{center}
\vskip 0.2in
\includegraphics[width=8cm,clip]{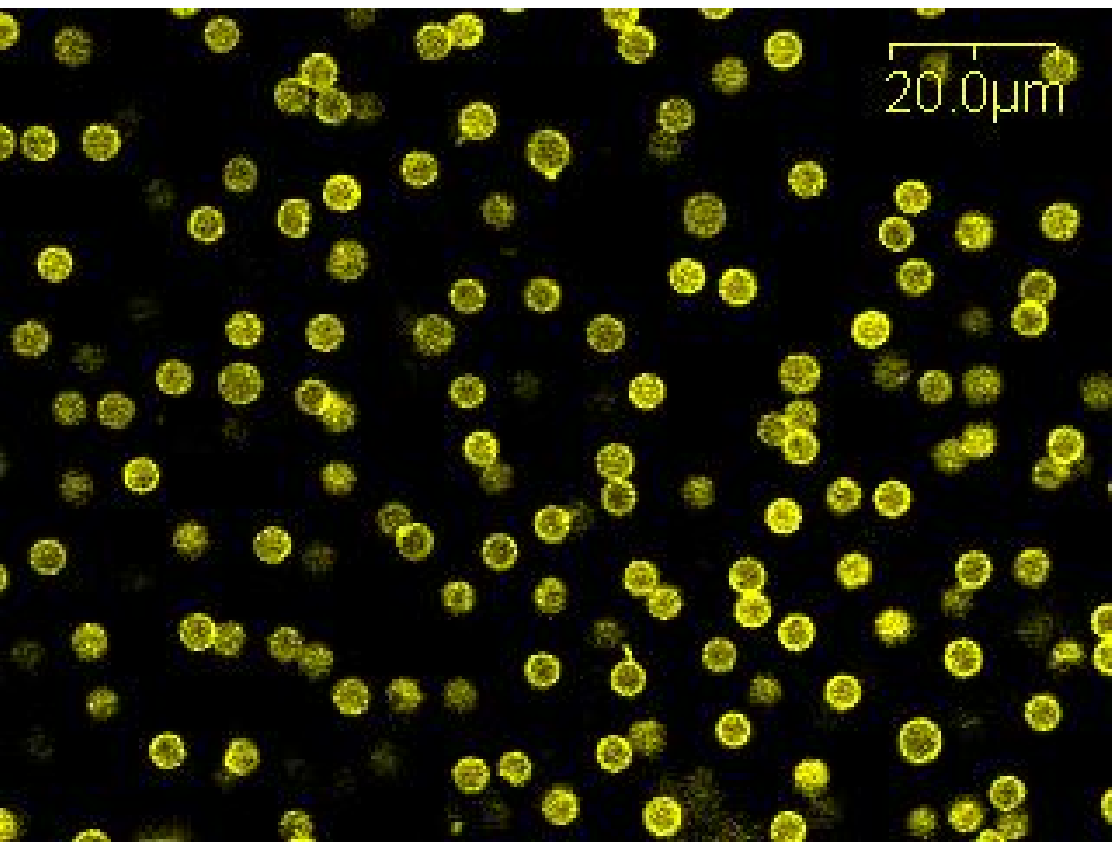}\\
\includegraphics[width=8cm,clip]{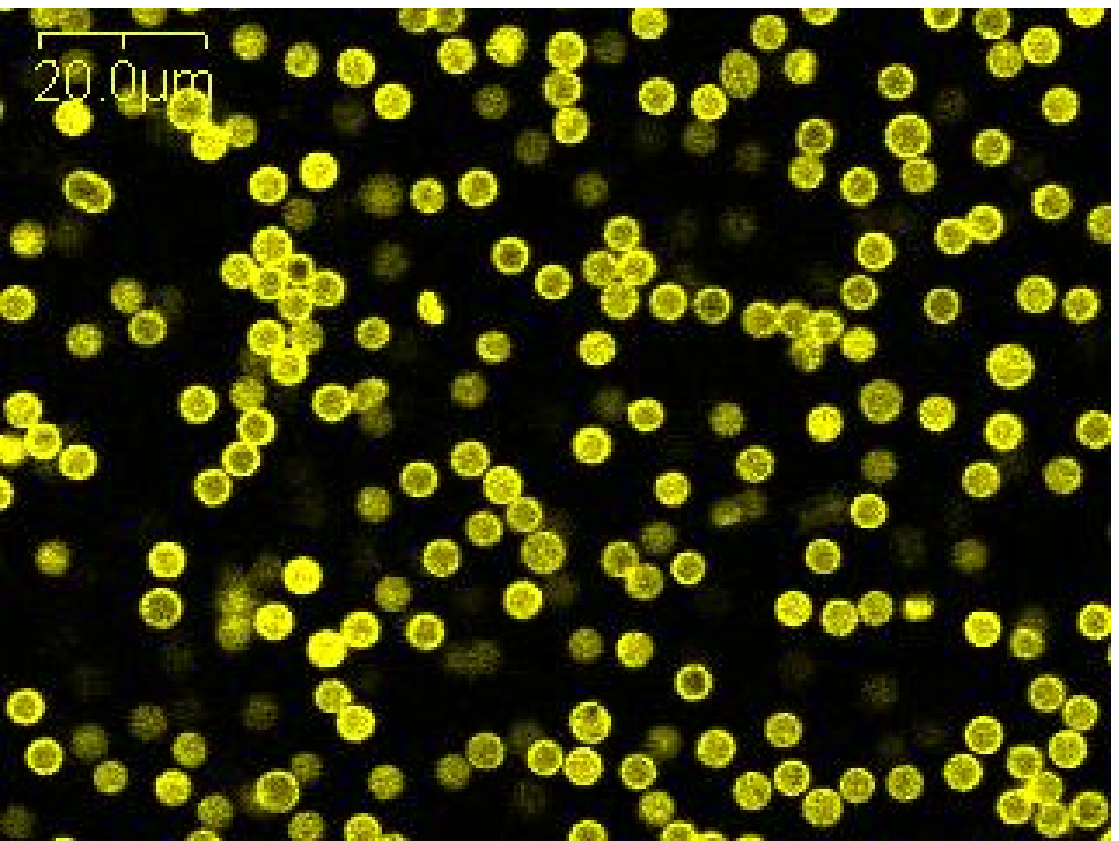}\\
\includegraphics[width=8cm,clip]{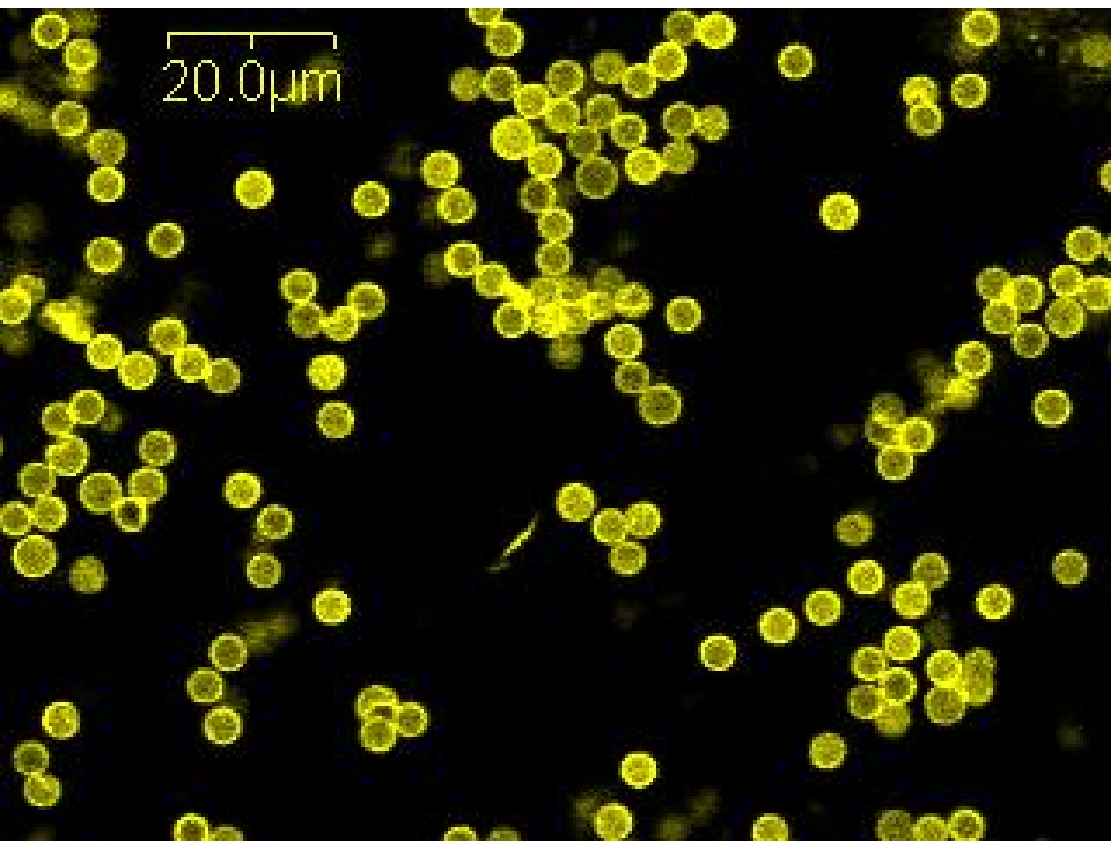}\\
\end{center}
\caption{Confocal images of the dispersion of capsules close to
the wall taken in one (top), four (middle), and five (bottom) days
after the preparation.
         }
\label{fig:2}
\end{figure}

Immediately after the preparation, the ``filled'' capsules have
been immersed into a glass bottom dish filled with water. Owing to
their greater density, the capsules slowly sediment and settle
near the bottom. Our first experiment, performed one day after
encapsulation, shows that the intimate contact between the
negatively charged wall (glass bottom) and the capsules is
prevented by a repulsive force. This is in evident contrast with
the behavior of positively charged original ``hollow'' capsules.
Strictly speaking, we were unable to perform the same type of
measurements with the ``hollow'' capsules. The difficulty is
related to that the sedimentation rate of ``hollow'' capsules is
much slower, so that they coagulate in the bulk before approaching
the wall. When these aggregates of ``hollow'' capsules approach
the wall, they attach the oppositely charged wall very quickly
without long-term stay in the vicinity of it. The behavior of
``filled'' capsules is also different from that of MF particles of
the same size, which were found to quickly adhere to the glass
surface. So, almost likely in case of ``filled'' capsules we
observe a double layer repulsion from the wall, which means that
``filled'' capsules also bear a negative charge. Since the
``filled'' capsules also undergo Brownian motion (as it was first
detected in~\cite{kim.bs:2004}), their location fluctuates about
an equilibrium height $L$ near the wall at which gravity and
double-layer repulsion are balanced. The confocal plane was then
chosen in such a distance from the wall, where the majority of
capsules were located (Fig.~\ref{fig:exam1_a}, top). It was then
found that capsules located in the confocal plane and exhibiting
normal fluctuations show random tangential motion (see inset in
Fig.~\ref{fig:1}). The normal and tangential components of the
Brownian fluctuations are affected differently by wall hindrance,
which is in agreement with hydrodynamic
theories~\cite{brenner.h:1961,goldman.aj:1967}. The typical
snapshot of the capsule configurations taken one day after
encapsulation is shown in Fig.~\ref{fig:2}, top. One can see that
capsules are isolated and move independently. An important point
to note is that the spatial distribution of capsules in the
confocal plane parallel to the wall is typical for a charge
stabilized dispersion. Further observations revealed the
development of the aggregation of capsules. In three days we have
observed a large amount of dimers and triplets (not shown), and
larger aggregates were formed in four days (Fig.~\ref{fig:2},
middle) after encapsulation. Analysis of a typical trajectory of
capsules suggest that, in average, they move slower and with the
smaller length of free-path (Fig.~\ref{fig:1}). After about one
week, however, most of capsules aggregate and sediment to the
wall, indicating that in case of large capsule aggregates
(Fig.~\ref{fig:2}, bottom) a double layer repulsion no longer
balances gravity. Capsules in aggregates are fully immobilize.

The electrostatic nature of coagulation slowing down has been
confirmed by investigating the behavior of a dispersion of
``filled'' microcapsules in the presence of salt ions. At low
concentration of NaCl (less than 0.1 mol/L), a dispersion of
``filled'' capsules coagulates very slowly, whereas at a higher
concentration of NaCl (more than 0.1 mol/L), where the
electrostatics are substantially screened, the aggregation
develops quite quickly.

After all ``filled'' capsules sedimented to the wall, we have
measured the inner distribution of PSS (Fig.~\ref{fig:exam1_a},
bottom). All measurements were done in water environment. We found
that this is non-uniform and reveals the concentration peaks of
the PSS at the capsule shell. The concentration of PSS near the
shell indicates that the interior of the shell has a
non-compensated negative charge, so that PSS ions tend to repel.
This seems to be a consequence of counter-ion
leakage~\cite{stukan.mr:2005}.

\subsection{Analysis}

The essential observations are these: the dispersion of ``filled''
(with strong polyanions) capsules shows behavior typical for
charge-stabilized colloids. Why, and what does it mean? For
dispersion of solid particles the slow rate of aggregation is
normally caused by an interplay of a long-range double-layer
repulsion and a shorter-range dispersion
attraction~\cite{derjaguin.bv:1941}. Therefore, a similar
coagulation behavior of the dispersion of microcapsules indicates
that we might deal with the same types of interaction forces.

Multilayer shells of capsules used here are positively charged.
The inner polyelectrolyte solution obeys the electro-neutrality
requirement, so that the total charge of counter-ions (Na$^+$ in
our case) is equal to a charge of dissociated polymer chains (i.e.
PSS$^-$). However, reasons for a negative charge of ``filled''
capsules are intuitively clear. It indicates a counterion leak-out
from the capsule interior~\cite{stukan.mr:2005}, which results
from an electro-chemical equilibrium between the outer solution
with higher entropy of the ions and the inner solution where the
counterions gain electrostatic energy. The fraction of the
escaping ions and, hence, the capsule charge should grow on
dilution of the dispersion and on lowering the ionic strength in
the solvent reservoir. This is consistent with the earlier results
confirming that a multilayer shell is a typical semi-permeable
membrane (see~\cite{vinogradova.oi:2004b,vinogradova.oi:2006} and
references therein) and with the non-uniform distribution of the
inner polyelectrolyte indicating the leakage of (positively
charged) counter-ions.

We suppose that the interaction potential for charged capsules
$U(r)$ has the DLVO form including the van der Waals and
electrostatic contributions
\begin{equation}
U(r) = U_{vdW}(r)+U_{el}(r) \label{potdlvo}
\end{equation}
For hollow capsules, however, the van der Waals term  $U_{vdW}(r)$
differs from the standard Hamaker potential for solid spheres,
which is normally applied. The corresponding interaction potential
for two shells of radius $R$ and thickness $h$ in the Derjaguin
approximation (the gap $d$ and the thickness are much smaller than
their radius $d,h \ll R$) can be conveniently written as a
function of the gap thickness $d$ \cite{tadmor.r:2001}
\begin{eqnarray}
U_{vdW}(d) &=& - \frac{A R}{12} \left(\frac{1}{d+2 h} - \frac{2}{d
+ h} + \frac{1}{d}\right)\nonumber\\ &-& \frac{A}{6} \ln\left(
\frac{d(d+2h)}{(d+h)^2} \right)
\end{eqnarray}
where $A$ is the Hamaker constant for the shell polyelectrolytes
in the given solvent. In the limit $d \ll h$, this potential can
be evaluated as $-A R/(12 d) \times 2h/(d + 2 h)$. So, at contact,
the dispersion force is the same as for a solid particle made of
the same material.

The electrostatic interaction of two charged spherical particles
is given by
\begin{equation}
U_{el}(r) = \frac{ \left( Z_{eff} e \right)^2}{4 \pi \varepsilon
\varepsilon_0} \left[ \frac{\exp(\kappa R)}{(1 + \kappa
R)}\right]^2 \frac{\exp(-\kappa r)}{r} \label{potel}
\end{equation}
where $Z_{eff}$ is the effective charge of the sphere, $e$ the
elementary charge, $\kappa^{-1}$ the Debye screening length,
$\varepsilon$ and $\varepsilon_0$ the absolute dielectric
permittivities of vacuum and the relative permittivity of the
solvent, respectively.

In a similar fashion, the interaction potential of the capsule
with the glass wall can be calculated, which can be used to
estimate the stability of the dispersion with respect to
deposition of the capsules on the cell walls. The interaction
potential for a thin spherical shell and a thick flat plate has a
form \cite{hamaker.hc:1937,tadmor.r:2001}
\begin{eqnarray}
U_{cw}(d) &=& \frac{A_{cw} R}{6} \left(\frac{1}{d} - \frac{1}{d+h}
\right) - \frac{A_{cw}}{6} \ln \left( \frac{d}{d+h} \right) +
\nonumber\\ &+& 64 \pi R \varepsilon \varepsilon_0 \gamma_1
\gamma_2 \left( \frac{k_B T}{e} \right)^2 \exp(-\kappa d)
\label{potwall}
\end{eqnarray}
where $A_{cw}$ is the Hamaker constant for glass-solvent-
polyelectrolyte materials combination, $\gamma_i=\tanh\left(e
\psi_{i}/(4 k_B T) \right)$, and $\psi_i$ the corresponding
effective surface potential of glass and capsule shell.

%

To characterize the capsule diffusion, we performed the analysis
of particle trajectories in each of the samples over 50 frames
using the particle tracking IDL scripts \cite{crocker.jc:1996}.
Our observations suggest that capsules can be conventionally
subdivided into two groups: slow capsules, with the net
displacement of $l<0.05 \mu m$ between the consequent frames, and
fast, with the average displacement of $0.2-0.4 \mu m$. The fast
capsules constitute the fraction of single particles, and the
decrease of this fraction with time can be used to evaluate the
capsule aggregation rate. Below we describe our approach. It
should be stressed, however, that our evaluation below should be
treated as orders of magnitude estimates. This is an inevitable
consequence of a limited experimental information we have.

We further consider only the \emph{suspended single particles},
which exhibit relatively fast Brownian motion, i.e. we excluded
short trajectories with the frame-to-frame displacement $l < 0.2
\mu m$ from the analysis. The self-diffusion coefficients
$D_{wall}$ were evaluated from the mean jump length, which is for
a two-dimensional random walk
$ \left\langle \Delta r^2 \right\rangle  = 4 D_{wall} \Delta t $
The observed capsule diffusion occurs close to the bottom of the
sample cell and therefore hindered due to hydrodynamic interaction
with the glass surface. A correction to the diffusion coefficient
near a wall is~\cite{Faxen}

\begin{equation}
\frac{D_{wall}}{D_0} = 1 - \frac{9}{16} \frac{R}{L} + \frac{1}{8}
\left( \frac{R}{L}\right)^3 -  \frac{45}{256} \left(\frac{R}{L}
\right)^4 - \frac{1}{16} \left( \frac{R}{L}\right)^5
\end{equation}
Here, $D_{wall}$ is the diffusion coefficient close to the stick
solid boundary (sample bottom).

\begin{figure}
\begin{center}
\vskip 0.2in
\includegraphics[width=8cm,clip]{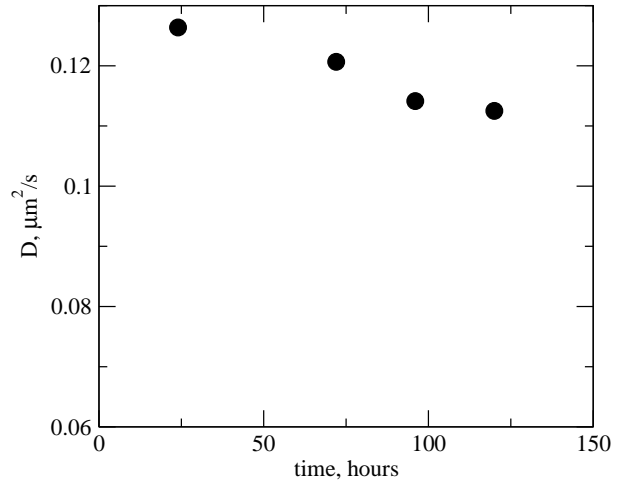}
\end{center}
\caption{ Lateral self-diffusion coefficient for filled capsules
corrected for the sphere-wall hydrodynamic interaction.
         }
\label{fig:4}
\end{figure}

From the observed diffusion coefficient, we can estimate the
distance to the glass wall. Assuming that the capsule
self-diffusion coefficient in the bulk is given by  the
Stokes-Einstein value $D_0 = k_B T / (6 \pi \eta R)$ (i.e.
neglecting the electrolyte friction effects), where $T$ is the
absolute temperature, $\eta$ the solvent viscosity. For a
spherical capsule of $4 \mu m$ in diameter $D_0$ amounts to $1.1
\times 10^{-13}$ m$^2$/s in an aqueous suspension at room
temperature. Then, using the Fax\'{e}n's formula for the 1-day old
sample, we find that the correction factor equals 0.47, i.e.
$D_{wall} = 0.47 D_0$, and, hence, the distance to the wall is of
the order of $d \approx 0.16 R \approx 300$ nm.


The data in Fig. \ref{fig:4} indicate some changes in the
diffusion coefficient with time. It is important to note that in
sedimentation process, the effective volume fraction of the
capsules close to the cell bottom (i.e. in the focal plane)
increases. We should therefore expect a decrease in the mobility
of the non-aggregated capsules. Indeed, in Fig. \ref{fig:4} the
diffusion constant of the free capsules is decreasing with the age
of the sample. Another observation that we can make out of the
image series is that the capsule aggregation happens easier than
their sedimentation and adhesion to the surface.

\begin{figure}
\begin{center}
\vskip 0.2in
\includegraphics[width=8cm,clip]{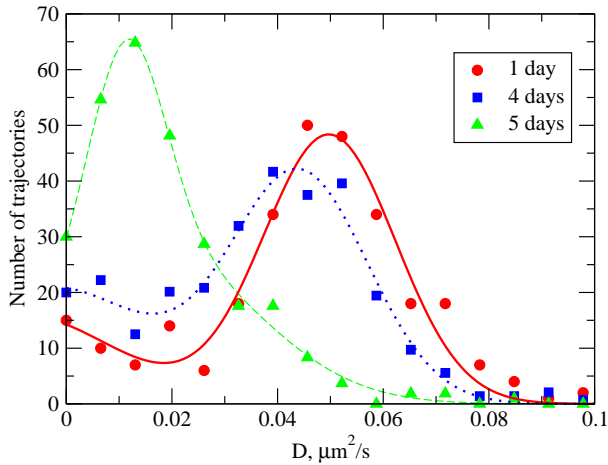}
\end{center}
\caption{Distributions of the self-diffusion coefficient for three
sample ages. The curves show the best fit to the histograms by a
sum of two Gaussian distributions.
         }
\label{fig:33}
\end{figure}

Now, we can estimate the effective charge and surface potential of
the filled capsules from the observed aggregation rate. The slow
aggregation we observe here is related to existence of the
potential barrier in the pair interaction between the capsules.
The aggregation rate of single particles is
\begin{equation}
\frac{dn}{dt} = - 16\pi R D n^2 W^{-1} \label{aggrate}
\end{equation}
where $n$ is the number density of single particles and $W$ the
stability ratio. In its turn, the latter can be related to the
pair interaction potential $U(r)$
\begin{equation}
W = 2R  \int_{2R}^\infty {\frac{ e^{U(r)/k_BT}}{r^2} dr}.
\label{stability}
\end{equation}
Using the model potential, Eq.~\ref{potdlvo}, with the Debye
length of 300 nm \cite{vinogradova.oi:2001}, the Hamaker constant
of polystyrene in water $A=1.3 \times 10^{-20}$J
\cite{viser.j:1976}, the shell thickness 10 nm, we evaluate the
stability ratio, from which we can extract the only remaining free
parameter, the capsule effective charge $Z_{eff}$.

\begin{figure}
\begin{center}
\vskip 0.2in
\includegraphics[width=8cm,clip]{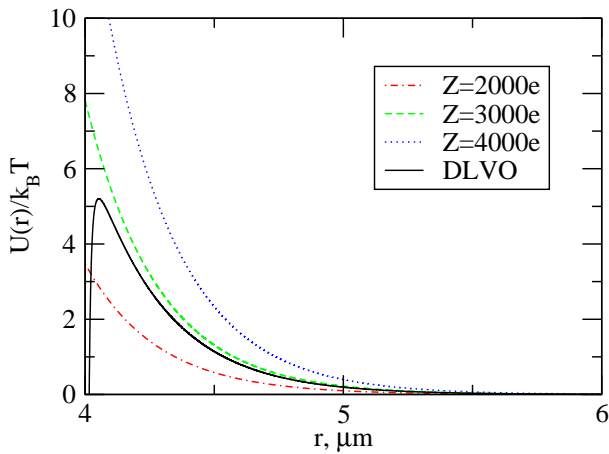}
\end{center}
\caption{ Effective pair potential for the filled capsules as
obtained from the analysis of the aggregation rate and capsule
diffusion from Eqs.~\ref{potdlvo}, \ref{aggrate}, and
\ref{stability}. Also shown are the curves for the electrostatic
part of the interaction, Eq.~\ref{potel} corresponding to
different effective charges of the capsule.
         }
\label{figpot}
\end{figure}

Now we substitute into Eq.~\ref{aggrate} the capsule
concentrations in the layer adjacent to the focal plane, near the
cell bottom. From the mean distance between the capsules and their
diameter we get the capsule volume fraction of 0.04. Then, by
estimating the initial aggregation rate from the fraction of free
capsules in the 1-day sample (0.7), we find $W=3400$. For fitting
the effective capsule charge, the potential, Eq.~\ref{potdlvo},
was substituted into integral, Eq.~\ref{stability}, and numerical
integration was performed. The best fit to the stability ratio
$W=3400$ was then obtained with $Z_{eff}=-2260$, which corresponds
roughly to the surface potential of -20 mV. Here we have to remind
that our multilayer shell is positively charged, so that the
negative potential we measure is solely due to encapsulated
polyelectrolyte. This potential calculated according to
Eq.~(\ref{potdlvo}) together with the curves for its electrostatic
part at $\left| Z_{eff} \right|=2000$, 3000, and 4000 are shown in
Fig.~\ref{figpot}. The very short-range part of the interaction
coming from the steric repulsion between the shells is not shown.
One can see that the filled capsules in a stabilized dispersion
would be separated by at least $1~\mu m$, the distance at which
the potential roughly equals $k_B T$. The above estimate of the
capsule charge must represent rather the lower boundary as we did
not account for attraction between the encapsulated material that
would make the van der Waals attraction term larger in magnitude.
Moreover, for the polyelectrolyte multilayers one can expect an
additional correlation term, which comes from the mutual
attraction of charge patches formed by polyelectrolytes adsorbed
at the surfaces of neighboring capsules, and is known to affect
the dispersion stability
\cite{gregory.jj:1973,miclavic.s:1994,bouyer.f:2001,grosberg.a:2002,lobaskin.v:2003}.
We should note that the charge $\left| Z_{eff} \right| =2260$ is
still much smaller than the charge saturation value for spherical
particles of $R=2 \mu m$, as can be estimated from the
Poisson-Boltzmann theory \cite{aubouy.m:2003}, which amounts to
90000. A dispersion of capsules can be therefore made more stable
by increasing the shell charge density by means of tuning the
chemical composition of the constituting polyelectrolytes or by
encapsulating a larger amount of polyelectrolytes. Clearly, the
stability with respect to capsule deposition on the glass walls
can be improved by changing the sign of the outermost
polyelectrolyte layer in the shell from polycation to polyanion.

\section{Conclusions}

In this work we report the dynamics and stability of a dispersion
of ``filled'' polyelectrolyte multilayer microcapsules as obtained
from the capsule aggregation kinetics, particle trajectories, and
spatial distributions. We observe that in contrast to the
dispersion of positively charged hollow capsules, which quickly
aggregate in the bulk and then stick to the glass bottom of the
sample cell, the capsules filled with strongly charged polyanions
remain suspended for several days, although they quickly sediment
and stay very close to the glass. We conclude that the net charge
of the filled capsules is not determined by the charge of the
multilayer shell anymore but is dominated by the charge of the
encapsulated material. We observe also that average mobility of
suspended capsules decreases as they sediment and their
aggregation occurs easier than their sedimentation and adhesion
onto the glass bottom of the sample cell.

\section*{Acknowledgements}

BSK acknowledges the financial support of the Alexander von
Humboldt Foundation at the initial stage of this study. We also
thank Eric Weeks for his kind assistance with using the particle
tracking routines.

\bibliography{kim6}

\end{document}